\documentclass[12pt, preprint]{aastex}

\shorttitle{\emph{Kepler}: oscillations in solar-type stars}
\shortauthors{Chaplin et al.}

\begin{document}

\title{The asteroseismic potential of \emph{Kepler}: first results for
 solar-type stars}

\author{
   W.~J.~Chaplin\altaffilmark{1},
   T.~Appourchaux\altaffilmark{2},
   Y.~Elsworth\altaffilmark{1},
   R.~A.~Garc\'ia\altaffilmark{3},
   G.~Houdek\altaffilmark{4},
   C.~Karoff\altaffilmark{1},
   T.~S.~Metcalfe\altaffilmark{5},
   J.~Molenda-\.Zakowicz\altaffilmark{6},
   M.~J.~P.~F.~G.~Monteiro\altaffilmark{7},
   M.~J.~Thompson\altaffilmark{8}
% ...
}
\altaffiltext{1}{School of Physics and Astronomy, University of Birmingham, Edgbaston, Birmingham, B15 2TT, UK}

\altaffiltext{2}{Institut d'Astrophysique Spatiale, Universit\'e Paris XI -- CNRS (UMR8617), Batiment 121, 91405 Orsay Cedex, France}

\altaffiltext{3}{Laboratoire AIM, CEA/DSM -- CNRS -- Universit\'e Paris Diderot -- IRFU/SAp, 91191 Gif-sur-Yvette Cedex, France}

\altaffiltext{4}{Institute of Astronomy, University of Vienna, A-1180 Vienna, Austria}

\altaffiltext{5}{High Altitude Observatory and, Scientific Computing Division, National Center for Atmospheric Research, Boulder, Colorado 80307, USA}

\altaffiltext{6}{Astronomical Institute, University of Wroc\l{}aw, ul. Kopernika, 11, 51-622 Wroc\l{}aw, Poland}

\altaffiltext{7}{Centro de Astrof\'\i sica, Universidade do Porto, Rua das Estrelas, 4150-762, Portugal}

\altaffiltext{8}{School of Mathematics and Statistics, University of Sheffield, Hounsfield Road, Sheffield S3 7RH, UK}

% ...

\begin{abstract}

We present preliminary asteroseismic results from \emph{Kepler} on
three G-type stars.  The observations, made at one-minute cadence
during the first 33.5\,d of science operations, reveal high
signal-to-noise solar-like oscillation spectra in all three stars:
About 20 modes of oscillation may be clearly distinguished in each
star. We discuss the appearance of the oscillation spectra, use the
frequencies and frequency separations to provide first results on the
radii, masses and ages of the stars, and comment in the light of these
results on prospects for inference on other solar-type stars that
\emph{Kepler} will observe.

\end{abstract}


\section{Introduction}
\label{sec:intro}

The \emph{Kepler} mission (Koch et al. 2010) will realize
significant advances in our understanding of stars, thanks to its
asteroseismology program, particularly for cool (solar-type)
main-sequence and subgiant stars that show solar-like oscillations,
i.e., small-amplitude oscillations intrinsically damped and
stochastically excited by the near-surface convection (e.g.,
Christensen-Dalsgaard 2004). Solar-like oscillation spectra have many
modes excited to observable amplitudes. The rich information content
of these seismic signatures means that the fundamental stellar
properties (e.g., mass, radius, and age) may be measured and the
internal structures constrained to levels that would not otherwise be
possible (e.g., see Gough 1987; Cunha et al. 2007).

High-precision results are presently available on a selection of
bright solar-type stars (e.g., see Aerts et al. 2010, and references
therein). Recent examples include studies of \emph{CoRoT} satellite
data (e.g., Michel et al. 2008; Appourchaux et al. 2008), and data
collected by episodic ground-based campaigns (e.g., Arentoft et
al. 2008). However, the \emph{Kepler} spacecraft will increase by more
than two orders of magnitude the number of stars for which
high-quality observations will be available, and will also provide
unprecedented multi-year observations of a selection of these stars.

During the initial ten months of science operations, \emph{Kepler}
will survey photometrically more than 1500 solar-type targets selected
for study by the \emph{Kepler Asteroseismic Science Consortium} (KASC)
(Gilliland et al.  2010a). Observations will be one month long for
each star. In order to aid preparations for analyses of these stars,
\emph{Kepler} data on three bright solar-type targets -- KIC~6603624,
KIC~3656476 and KIC~11026764 -- have been made available in a
preliminary release to KASC (see Table~1 for the 2MASS ID of each
star).  The stars are at the bright end of the \emph{Kepler} target
range, having apparent magnitudes of 9.1, 9.5 and 9.3\,mag,
respectively.  The \emph{Kepler Input Catalog} (KIC; Latham et
al. 2005), from which all KASC targets were selected, categorizes them
as G-type stars. In this \emph{Letter} we present initial results on
these stars.

\section{The Frequency Power Spectra of the Stars}
\label{sec:pow}

\begin{figure*}
\epsscale{1.0}
\plotone{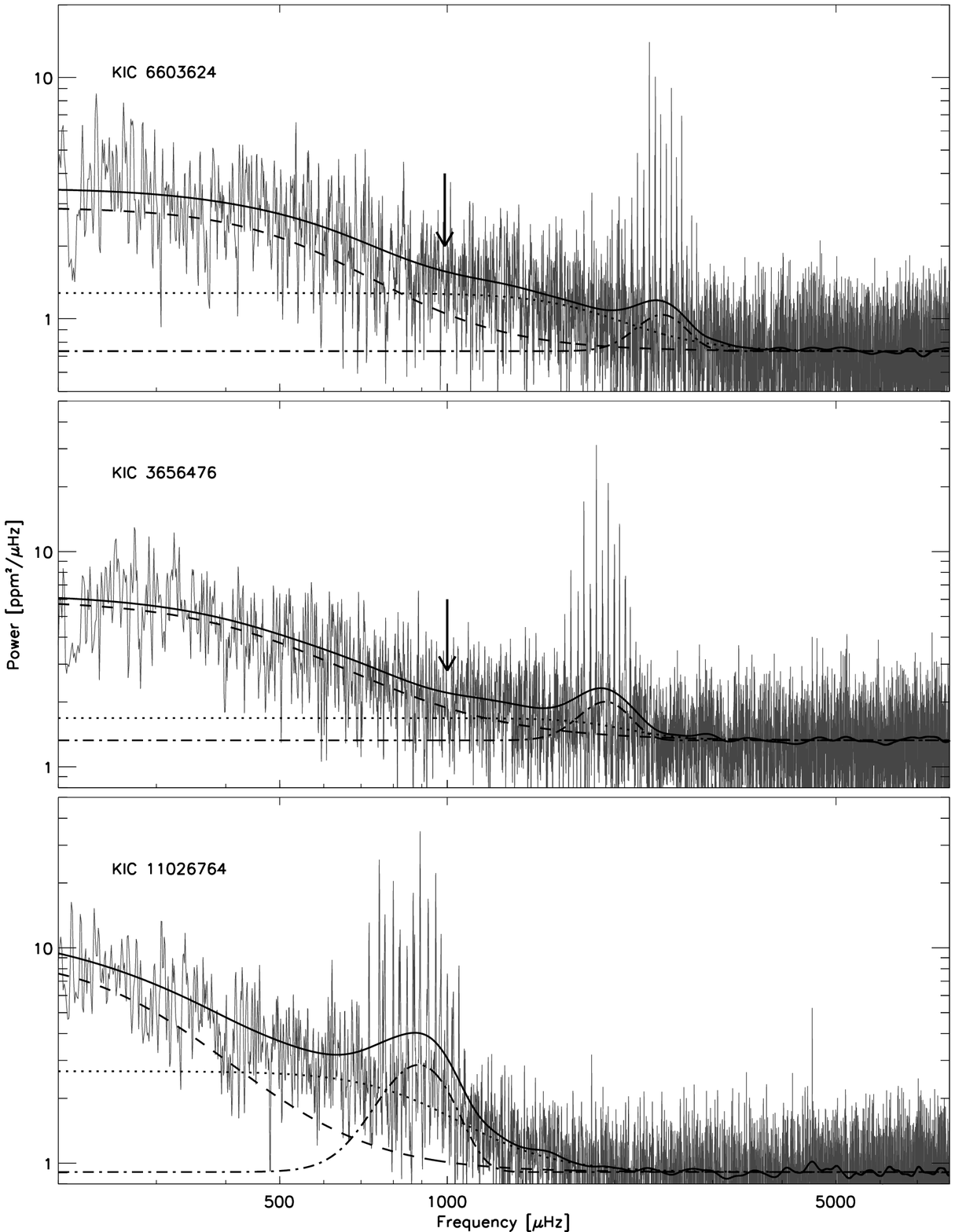}
\caption{Frequency-power spectra of the three stars, smoothed over
$1\,\rm \mu Hz$, plotted on a log-log scale (grey). The continuous
black lines show heavily smoothed spectra (Gaussian filter). The other
lines show best-fitting estimates of different components: the power
envelope due to oscillations, added to the offset from shot noise
(dot-dashed); faculae (dotted) and granulation (dashed). The arrows
indicate kinks in the rising background (see text).}

\label{fig:specs}
\end{figure*}

The stars were observed for the first 33.5\,days of science operations
(2009 May 12 to June 14). Detection of oscillations of cool
main-sequence and subgiant stars demands use of the 58.85-sec,
high-cadence observations since the dominant periods in some
solar-type stars can be as short as two minutes.  Timeseries data were
then prepared from the raw observations in the manner described by
Gilliland et al. (2010b). Fig.~\ref{fig:specs} shows the
frequency-power spectra on a log-log scale (grey). Heavily smoothed
spectra (Gaussian filter) are over-plotted as continuous black
lines. The quality of these data for asteroseismology is excellent,
with each spectrum showing a clear excess of power due to solar-like
oscillations. The excess is in each case imposed upon a background
that rises slowly toward lower frequencies.

 \subsection{Power Spectral Density of the Background}
 \label{sec:bg}

The average power spectral density at high frequencies provides a good
measure of the power due to photon shot noise. The high-frequency
power is in all cases close to pre-launch estimates, given the
apparent magnitudes of the targets, suggesting that the data are close
to being shot-noise limited (e.g., see Gilliland et al. 2010b). This
result lends further confidence to our expectations -- from
hare-and-hounds exercises with simulated data -- that we will be able
to conduct asteroseismology of solar-like KASC survey targets down to
apparent magnitudes of 11 and fainter (e.g., see Stello et al. 2009).

There are also background components of stellar origin, and we
describe these components by power laws in frequency.  Kinks in the
observed backgrounds of KIC~6603624 and KIC~3656476 (kinks in solid
black curves, indicated by arrows on Fig.~\ref{fig:specs}) suggest the
presence of two stellar components in the plotted frequency range. One
component (dotted lines) is possibly the signature of bright
faculae. A similar signature, due to faculae, is seen clearly in
frequency-power spectra of Sun-as-a-star data collected by the
\emph{VIRGO/SPM} photometers on \emph{SOHO} (e.g., Aigrain et
al. 2004). We do not see a kink in the background of KIC~11026764, and
it may be that the characteristic ``knee'' of the facular component
coincides in frequency with the oscillation envelope, making it hard
to discriminate from the other components. The other stellar
component, which is shown by all three stars (dashed lines), carries
the signature of stellar granulation.

 \subsection{The Oscillation Spectra}
 \label{sec:osc}

\begin{figure*}
\epsscale{1.0}
\plotone{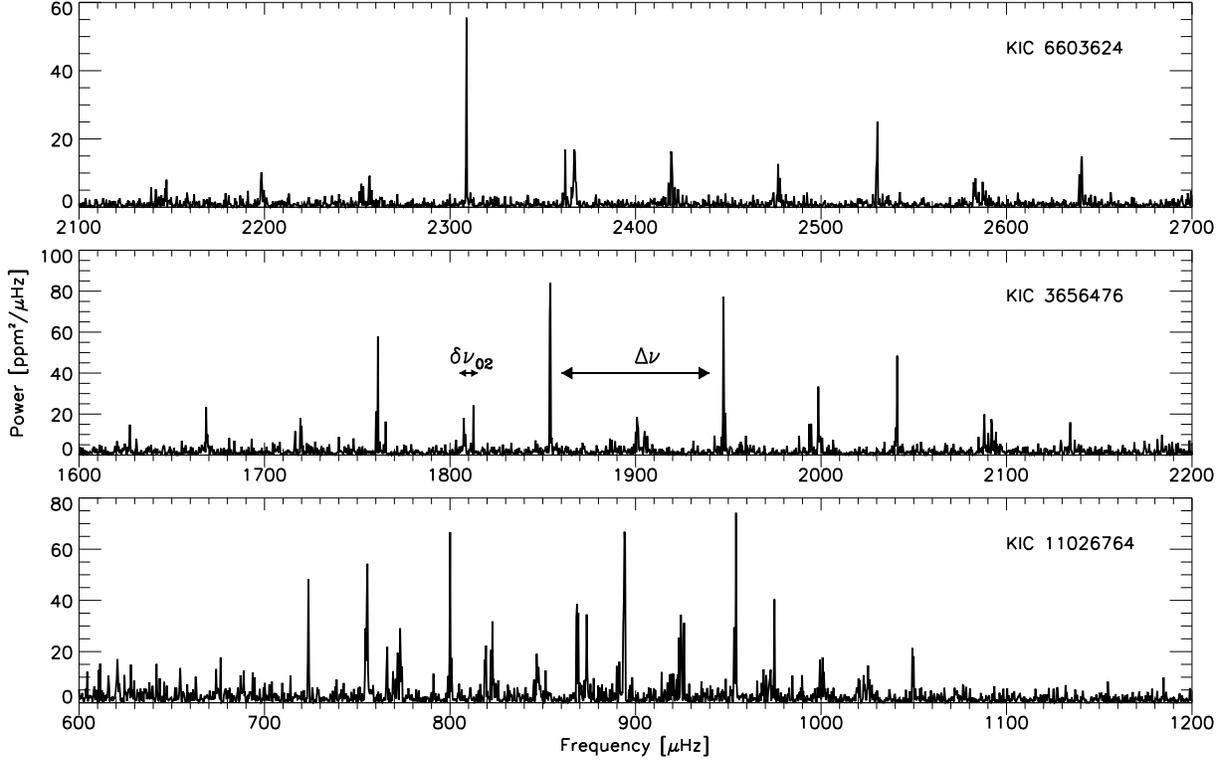}
\caption{Frequency-power spectra of the three stars, plotted on a
  linear scale over the frequency ranges where the mode amplitudes are
  most prominent. Examples of the characteristic large ($\Delta\nu$)
  and small ($\delta\nu_{02}$) frequency separations are also marked on
  the spectrum of KIC~3656476.}
\label{fig:zoom}
\end{figure*}

Fig.~\ref{fig:zoom} shows in more detail the oscillation spectra of
the three stars. The high signal-to-noise ratios observed in the mode
peaks allows about 20 individual modes to be identified very clearly
in each star.

Stars KIC~6603624 and KIC~3656476 present patterns of peaks that all
show nearly regular spacings in frequency. These peaks are due to
acoustic (pressure, or p) modes of high radial order, $n$, with
frequencies approximately proportional to $\sqrt{\bar \rho}$, $\bar
\rho \propto M/R^3$ being the mean density of the star, with mass $M$
and surface radius $R$. The most obvious spacings are the ``large
frequency separations'', $\Delta\nu$, between consecutive overtones of
the same spherical angular degree, $l$. These large separations are
related to the acoustic radii of the stars. The ``small frequency
separations'' are the spacings between modes adjacent in frequency
that have the same parity angular degree. Here, we see clearly the
small separations $\delta\nu_{02}$ between modes of $l=0$ and $l=2$
(photometric observations have low sensitivity to $l=3$ modes, and so
they cannot be seen clearly in these frequency-power spectra). The
small separations are very sensitive to the gradient of the sound
speed in the stellar cores and hence the evolutionary state of the
stars.

The near regularity of the frequency separations of KIC~6603624 and
KIC~3656476 is displayed in the \'echelle diagrams in
Fig.~\ref{fig:echelle}. Here, we have plotted estimates of the mode
frequencies of the stars (see Section~\ref{sec:inf} below) against
those frequencies modulo the average large frequency separations. In a
simple asymptotic description of high-order p modes (e.g., Tassoul
1980) the various separations do not change with frequency. Stars that
obeyed this description would show vertical, straight ridges in the
\'echelle diagram (assuming use of the correct
$\Delta\nu$). Solar-type stars do in practice show departures of
varying degrees from this simple description. These variations with
frequency carry signatures of, for example, regions of abrupt
structural change in the stellar interiors, e.g., the near-surface
ionization zones and the bases of the convective envelopes (Houdek \&
Gough 2007).

While the variations are clearly modest in KIC~6603624 and
KIC~3656476, KIC~11026764 presents a somewhat different picture. Its
$l=1$ ridge is noticeably disrupted, and shows clear evidence of
so-called avoided crossings (Osaki 1975; Aizenman et al. 1977), where
modes resonating in different cavities exist at virtually the same
frequency.

These avoided crossings are a tell-tale indicator that the star has
evolved significantly. In young solar-type stars there is a clear
distinction between the frequency ranges that will support p modes and
buoyancy (gravity, or g) modes.  As stars evolve, the maximum buoyancy
(Brunt-V\"ais\"al\"a) frequency increases. After exhaustion of the
central hydrogen, the buoyancy frequency in the deep stellar interior
may increase to such an extent that it extends into the frequency
range of the high-order acoustic modes. Interactions between acoustic
modes and buoyancy modes may then lead to a series of avoided
crossings, which affect (or ``bump'') the frequencies and also change
the intrinsic properties of the modes, with some taking on mixed p and
g characteristics. The jagged appearance of the $l=1$ ridge of
KIC~11026764 illustrates these effects, which are strikingly similar
to those seen in ground-based asteroseismic data on the bright stars
$\eta$\,Boo (Christensen-Dalsgaard et al. 1995; Kjeldsen et al. 1995)
and $\beta$~Hyi (Bedding et al. 2007). Deheuvels \& Michel (2009) and
Deheuvels et al. (2009) have also reported evidence for avoided
crossings in CoRoT observations of the oscillation spectrum of the
star HD49385.

Next, we consider the peaks of individual modes. The observed
oscillation modes in solar-type stars are intrinsically stable
(e.g. Balmforth 1992a; Houdek et al. 1999) but driven stochastically
by the vigorous turbulence in the superficial stellar layers
(e.g. Goldreich \& Keeley 1977; Balmforth 1992b; Samadi \& Goupil
2001).  Solar-like mode peaks have an underlying form that follows, to
a reasonable approximation, a Lorentzian. The widths of the
Lorentzians provide a measure of the linear damping rates, while the
amplitudes are determined by the delicate balance between the
excitation and damping. Measurement of these parameters therefore
provides the means to infer various important properties of the still
poorly understood near-surface convection.

The observed maximum mode amplitudes are all higher than solar. This
is in line with predictions from simple scaling relations (Kjeldsen \&
Bedding 1995; Samadi et al. 2007), which use the inferred fundamental
stellar properties (see Section~\ref{sec:inf} below) as input. Data
from a larger selection of survey stars are required before we can say
anything more definitive about the relations.

It is clear even from simple visual inspection of the spectra that the
intrinsic linewidths of the most prominent modes are comparable in
size in all three stars to the linewidths shown by the most prominent
solar p modes ($\approx 1\,\rm \mu Hz$). It would not otherwise be
possible to distinguish the $l=0$ and $l=2$ modes so easily. The
intrinsic frequency resolution of these short survey spectra ($\sim
0.35\,\rm \mu Hz$) makes it difficult to provide more definitive
widths. However, the appearance of the modes in these stars is
consistent with the suggestion of Chaplin et al. (2009) that the
linewidths are a strong function of effective temperature. The three
stars here all have effective temperatures that are similar to, or
slightly cooler than, solar; while the F-type main-sequence stars
observed by \emph{CoRoT} -- which have effective temperatures a few
hundred degrees hotter than the Sun -- exhibit linewidths that are
several times larger than solar (e.g., see Appourchaux et al. 2008;
Barban et al. 2009; Garc\'ia et al. 2009). (We add that CoRoT sees
linewidths in the G-type star HD49385 (Deheuvels et al. 2009) that are
also narrower than those observed in F-type stars.)

Given the inference on the linewidths of the stars, we are able to
state with renewed confidence that it should be possible to extract
accurate and precise frequencies of $l=0$, 1 \emph{and} 2 modes in
many of the brighter solar-type KASC targets observed by
\emph{Kepler}, because the peaks will be clearly distinguishable. The
prospects on cool stars selected for long-term observations are
particularly good. Here, the combination of improved resolution in
frequency and modest (i.e., solar-like) linewidths will in principle
allow for robust estimation of frequency splittings of modes. These
splittings have contributions from the internal stellar rotation and
magnetic fields. We add that longer-term observations of the three
survey stars reported in this \emph{Letter} will be needed to show
indisputable evidence of frequency splitting.

 \section{Inference on the Stellar Properties}
 \label{sec:inf}

An estimate of the average separations $\Delta\nu$ and
$\delta\nu_{02}$ provides a complementary set of seismic data well
suited to constraining the stellar properties (Christensen-Dalsgaard
1993). In faint KASC survey targets -- where lower signal-to-noise
ratios will make it difficult to extract robust estimates of
individual frequencies -- the average separations will be the primary
seismic input data. The signatures of these separations are quite
amenable to extraction, owing to their near-regularity. 

\begin{figure*}
\epsscale{1.0}
\plotone{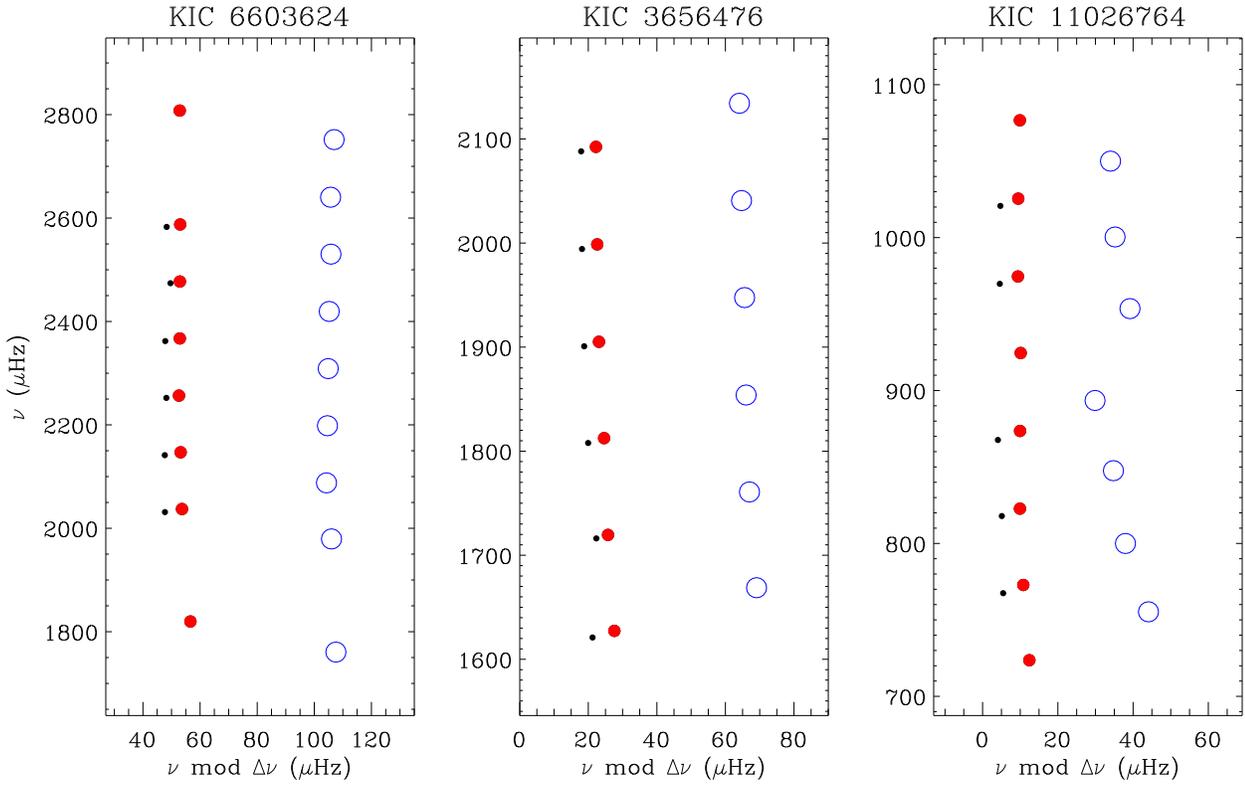}
\caption{\'Echelle diagrams of the observed frequencies in each star,
  showing the $l=0$ (filled red symbols), $l=1$ (open blue symbols)
  and $l=2$ (small black symbols) ridges.}
\label{fig:echelle}
\end{figure*}

Different teams extracted estimates of the average separations of the
three stars, with analysis methods based largely on use of the
autocorrelation of either the timeseries or the power spectrum (e.g.,
see Huber et al. 2009; Mosser \& Appourchaux 2009; Roxburgh 2009;
Hekker et al. 2010; Mathur et al. 2010). We found good agreement
between the different estimates (i.e., at the level of precision of
the quoted parameter uncertainties). Representative estimates of the
separations are given in Table~1. The teams also used peak-fitting
techniques (like those applied to CoRoT data; see, for example,
Appourchaux et al. 2008) to make available to the modeling teams
initial estimates of the individual mode frequencies. The \'echelle
diagrams plotted in Fig.~\ref{fig:echelle} show representative sets of
these frequencies.

Several modeling teams then applied codes to estimate the stellar
properties using the frequency separations, and other non-seismic data
(see below), as input; the results from these analyses were then used
as the starting points for further modeling, involving comparisons of
the observed frequencies with frequencies calculated from evolutionary
models. Use of individual frequencies increases the information
content provided by the seismic data (e.g., see Monteiro et al. 2000;
Roxburgh \& Vorontsov 2003; Mazumdar et al. 2006; Cunha \& Metcalfe
2007). (For a general discussion of the modeling methods, see Cunha et
al. 2007; Stello et al. 2009; Aerts et al. 2010; and references
therein. Further detailed presentations of the modeling techniques
applied here will appear in future papers.)

The frequencies and the frequency separations depend to some extent on
the detailed physics assumed in the stellar models (Monteiro et
al. 2002). Consequently, a more secure determination of the stellar
properties is possible when other complementary information -- such as
effective temperature $T_{\rm eff}$, luminosity (or $\log g$) and
metallicity -- is known to sufficiently high accuracy and
precision. The potential to test the input physics of models of field
stars (e.g., convective energy transport, diffusion, opacities, etc.)
requires non-seismic data for the seismic diagnostics to be effective
(e.g. see Creevey et al. 2007). The modeling analyses therefore also
incorporated non-seismic constraints, using $T_{\rm eff}$, metallicity
([Fe/H]) and $\log g$ from complementary ground-based spectroscopic
observations (see Table~1).  The uncertainties in these
spectroscopically estimated parameters are significantly lower than
those given by the KIC parameters. Preliminary results given by
different groups on the same ground-based spectra of these stars do
however suggest that the true, external errors are higher than the
quoted errors. Follow-up spectroscopic and photometric observations,
and further comparative analyses, are now being carried out for other
bright solar-type KASC targets by more than twenty members of KASC
(e.g., Molenda-\.Zakowicz et al. 2008).

Our initial estimates of the stellar radii and masses are presented in
Table~1. Given the close relation between the global properties of the
stars and their oscillation frequencies, these seismically inferred
properties are more precise, and more accurate, than properties
inferred without the seismic inputs. The radii of KIC~6603624 and
KIC~3656476 have been determined to better than 2\,\%, and the masses
to better than 6\,\%. The modeling suggests both stars may be near the
end of their main-sequence lifetimes.

The precision achieved for KIC~11026764 is not quite as good (about
5\,\% in the radius, and about 10\,\% in the mass). This star has
evolved off the main sequence, and is harder to model. KIC~11026764
demonstrates that when mixed modes are observed individual frequencies
can provide more stringent tests of the modeling than can the average
separations.  Initial results from modeling individual frequencies
show it is possible to reproduce the disrupted $l=1$ frequency ridge
(Fig.~\ref{fig:echelle}), indicating an age in the range 6 and
7\,Gyr. The modeling also suggests that some of the $l=2$ modes may
have mixed character.

Initial modeling results may also help to interpret the observed
seismic spectra, allowing further mode frequencies to be identified
securely. A possible example for KIC~11026764 concerns the prominent
mode at $\sim 723\,\rm \mu Hz$. The peak lies on the $l=0$ ridge, yet
its appearance -- a narrow peak, indicating a lightly damped mode
having mixed characteristics -- suggests a possible alternative
explanation: the modeling points strongly to the observed power being
predominantly from an $l=1$ mode that has been shifted so far in
frequency that it lies on top of an $l=0$ mode.

In summary, all three stars are clearly excellent candidates for
long-term observations by \emph{Kepler}. With up to 1\,yr of data we
would, for example, expect to measure the depths of the near-surface
convection zones, and the signatures of near-surface ionization of
He. It should also be possible to constrain the rotational frequency
splittings. With 2\,yr of data and more, we would also hope to begin
to constrain any long-term changes to the frequencies and other mode
parameters due to stellar-cycle effects (Karoff et al. 2009). More
detailed modeling will also allow us to characterize the functional
form of any required near-surface corrections to the model frequencies
(see Kjeldsen et al.~2008).

\acknowledgements Funding for this Discovery mission is provided by
NASA's Science Mission Directorate. The authors wish to thank the
entire \emph{Kepler} team, without whom these results would not be
possible.  We also thank all funding councils and agencies that have
supported the activities of KASC Working Group\,1, and the
International Space Science Institute (ISSI). The analyses reported in
this Letter also used observations made with FIES at the Nordic
Optical Telescope, and with SOPHIE at Observatoire de Haute-Provence.

{\it Facilities:} \facility{The Kepler Mission}

\begin{deluxetable}{lccccccccccc}
\tabletypesize{\scriptsize} \tablecaption{Non-seismic and seismic parameters, and preliminary stellar properties\tablenotemark{a}} 
\tablehead{
\colhead{Star}& \colhead{2MASS}& \colhead{$T_{\rm eff}$}& \colhead{log
$g$} & \colhead{[Fe/H]} & \colhead{$\Delta \nu$} & \colhead{$\delta
\nu_{02}$} & \colhead{$R$} & \colhead{$M$}\\ \colhead{} & \colhead{ID} &
\colhead{[K]} & \colhead{[dex]} & \colhead{[dex]} &
\colhead{[$\mu$Hz]} & \colhead{[$\mu$Hz]} & \colhead{[R$_{\sun}$]} &
\colhead{[M$_{\sun}$]} }

\startdata
KIC 6603624\tablenotemark{b}& 19241119+4203097& 5790$\pm$100& 4.56$\pm$0.10& 0.38$\pm$0.09&110.2$\pm$0.6& 4.7$\pm$0.2& 1.18$\pm$0.02& 1.05$\pm$0.06\\
KIC 3656476\tablenotemark{c}& 19364879+3842568& 5666$\pm$100& 4.32$\pm$0.06& 0.22$\pm$0.04& 94.1$\pm$0.6& 4.4$\pm$0.2& 1.31$\pm$0.02& 1.04$\pm$0.06\\
KIC 11026764\tablenotemark{b}& 19212465+4830532& 5640$\pm$80& 3.84$\pm$0.10& 0.02$\pm$0.06& 50.8$\pm$0.3& 4.3$\pm$0.5& 2.10$\pm$0.10& 1.10$\pm$0.12\\

\enddata

\tablenotetext{a}{Non-seismic parameters are $T_{\rm eff}$, log$g$ and
[Fe/H]; seismic parameters are $\Delta \nu$ and $\delta \nu_{02}$; and
the stellar properties inferred from the non-seismic and seismic data
are $M$ and $R$.}

\tablenotetext{b}{Non-seismic parameters for KIC~6603624 and
KIC~11026764 from observations made with FIES at the Nordic Optical
Telescope (NOT); data reduced in the manner of Bruntt et al. (2004,
2008).}

\tablenotetext{c}{Non-seismic parameters for KIC~3656476 from
observations made with SOPHIE at Observatoire de Haute-Provence; data
reduced in the manner of Santos et al. (2004).}

\label{tab:res}
\end{deluxetable}

\end{document}